# A P2P Network Topology for Optimizing Partition Tolerance to Reach the CAP Guarantee Bound in Consortium Blockchains

Han Wang, Hui Li, Zixian Wang, Baofu Han, Minglong Zhang,
Peter Han Joo Chong, *Senior Member, IEEE,* Xiaoli Chu, *Senior Member, IEEE,*
Yang Liu, Soung-Yue Liew, and Lunchakorn Wuttisittikulkij

**Abstract**—Decentralized cryptocurrency systems, known as blockchains, have shown promise as infrastructure for mutually distrustful parties to agree on transactions safely. However, Bitcoin-derived blockchains and their variants suffer from the limitations of the CAP Trilemma, which is difficult to be solved through optimizing consensus protocols. Moreover, the P2P network of blockchains have problems in efficiency and reliability without considering the matching of physical and logical topologies. For the CAP Trilemma in consortium blockchains, we propose a physical topology based on the multi-dimensional hypercube with excellent partition tolerance in probability. On the other hand, the general hypercube has advantages in solving the mismatch problems in P2P networks. The general topology is further extended to a hierarchical recursive topology with more medium links or short links to balance the reliability requirements and the costs of physical network. We prove that the hypercube topology has better partition tolerance than the regular rooted tree topology and the ring lattice topology, and effectively fits the upper-layer protocols. As a result, blockchains constructed by the proposed topology can reach the CAP guarantee bound through adopting the proper transmission and the consensus protocols protocols that have strong consistency and availability.

✦

## 1 INTRODUCTION

BITCOIN [1], as the first widely-deployed, decentralized global cryptocurrency, has sparked hundreds of variant systems. The core technological innovation powering these systems is the decentralized infrastructure known as the blockchain. Although numerous protocols have been proposed to improve the performance of blockchains, none of them get rid of the CAP (Consistency, Availability, and Partition Tolerance) Trilemma [2] in the distributed theory. Specifically, consistency, availability, and partition tolerance cannot be strongly satisfied simultaneously in the distributed system.

In order to balance the three properties in CAP Trilemma, some approaches optimize at the consensus layer. Original Nakamoto consensus meet good partition tolerance through the longest chain rule to choose the main chain. However, forks in the main chain would turn the strong consistency to the final consistency within a period, which results in long confirming time and limited throughput. In public blockchains, several alternative ledger structures [26], [27], [28], [29], [30] are explored to make the most of computing power. The performance of these protocols is still limited by the network synchronization rate because of their weak or final consistency. Once blocks are generated faster than the network synchronization rate, the number of forks would expand and the system would be insecure. In consortium blockchains and private blockchains, BFT (Byzantine Fault Tolerant) consensus protocols such as PBFT [21] are widely used to avoid forks [23]. Although nodes can reach agreement before the next round of consensus, BFT-like protocols have the disadvangates of communication complexity and scalability. Moreover, some performance-driven consensus protocols [3], [4], [5], [6], [24], [25] guarantee consistency and partition tolerance by delegation mechanism but decrease the number of core nodes participating in consensus. A. Lewis-Pye and T. Roughgarden [22] have further

- *Han Wang, Hui Li, Zixian Wang, and Baofu Han are with the Shenzhen Graduate School, Peking University, Shenzhen, China. E-mail: {wanghan2017, wangzixian}@pku.edu.cn, lih64@pkusz.edu.cn, Baofu-Han@hotmail.com.*
- *Minglong Zhang and Peter Han Joo Chong are with the Department of Electrical and Electronic Engineering, Auckland University of Technology, Auckland, New Zealand. E-mail: {mizhang, peter.chong}@aut.ac.nz.*
- *Xiaoli Chu and Yang Liu are with the Department of Electronic and Electrical Engineering, The University of Sheffield, Sheffield, U.K. E-mail: {x.chu, yangliu}@sheffield.ac.uk.*
- *Soung-Yue Liew is with the Faculty of Information and Communication Technology, Universiti Tunku Abdul Rahman, Kampar, Malaysia. E-mail: syliew@utar.edu.my.*
- *Lunchakorn Wuttisittikulkij is with Wireless Communication Ecosystem Research Unit, the Department of Electrical Engineering, Chulalongkorn University, Bangkok, Thailand. E-mail: wlunchak@chula.ac.th.*

proved an analog of the CAP Trilemma at the consensus layer that no protocol is both adaptive and has finality in the unsized and partially synchronous network.

In addition to optimizing consensus protocols, another idea is to optimize the P2P (Peer-to-Peer) network [10]. Many consortium blockchains use the structured P2P network without considering the matching of the physical topology. As a kind of overlay network, the query of P2P network consists of logical hops. According to the routing information saved by each node, a path with the least logical hops is selected as the optimal query path. However, the two nodes with close logic distance in the overlay network are often not close at the physical layer, which affects the query efficiency and reliability.

In this paper, we construct an extensible physical topology with excellent partition tolerance in probability to reduce the limitations of the upper-layer blockchain protocols. The proposed general physical topology is based on the multi-dimensional hypercube. It can alleviate the topological mismatch of structured P2P networks. We then design the hierarchical recursive physical topology to make the approach implementable. Our experiments show that the proposed physical topology does not affect the performance of the upper-layer protocol. As a result, our approach ensures good partition tolerance. Combined with upper-layer protocols having good consistency and availability, such as the PPoV (Parallel Proof of Vote) consensus protocol [11], the system can reach CAP boundaries.

**Roadmap.** In Section 2, we introduce the P2P network used in blockchains. Section 3 proposes the general and recursive topology construction methods based on hypercubes. We discuss the partition tolerance property and the link consumption in Section 4. Then we deploy the proposed topology and evaluate the consistency and availability performance in Section 5. Finally, we conclude our work in Section 6.

## 2 RELATED WORK

In the P2P network, each node is both a server and a client. Message exchange relies on the client group rather than the central server, so the nodes should participate in the relay. The realization of services is carried out directly between nodes without the intervention of intermediate nodes. As the service is distributed among nodes, the failure of some nodes or links has little impact on other parts of the network. On the other hand, the intermediate cost of exchanging message is low, which reduces the resource and time cost caused by centralization.

According to whether the network is centralized, the P2P network is divided into fully distributed P2P network and semi-distributed P2P network.

### 2.1 Fully Distributed P2P Network

The fully distributed P2P network includes structured and unstructured networks. The difference is whether the node address is structured. Nodes in the fully distributed P2P network can join and exit freely, and there is not any central node. In the unstructed P2P network, the entire network structure is a random graph structure without fixed network topologies and structured unified node addresses. The typical blockchain application of a fully distributed unstructured P2P network is Bitcoin. On the contrary, the structured P2P network defines the topological relationship of network nodes, and the topology structure of the network is assured by certain protocols between nodes. The typical blockchain application for fully distributed structured P2P networks is Ethereum [7].

### 2.2 Semi-Distributed P2P Network

The semi-distributed P2P network combines the advantages of structured and unstructured networks. The advantage of the semi-distributed P2P network is that it has good efficiency and scalability, and is easy to manage. Its typical blockchain application is Hyperledger [13].

Semi-distributed P2P networks are often combined with the delegated mechanism in the blockchain consensus algorithm. Specifically, according to evaluation criteria such as computing power, bandwidth, and retention time, it divides nodes into super nodes and ordinary nodes. The super node endorses and supervises the ordinary nodes of the organization or institution to which the node belongs, and participates in the core consensus process. Ordinary nodes make transactions through super nodes but do not participate in bookkeeping. Super nodes are equal to each other, and there is no frequent joining and exiting problem. Meanwhile, the number of super nodes in the consortium blockchain is far less than the number of nodes in the public blockchain. Although the original Bitcoin is based on the unstructured and fully distributed P2P network and flood mechanism, from the perspective of improving network efficiency, the semi-distributed P2P network is more suitable as the networking mode of the consortium blockchain.

### 2.3 P2P Network Topology in Blockchains

There has been some research focused on optimizing the topology of the structured P2P network to speed up the propagation of transactions and blocks in blockchains. C. Decker and R. Wattenhofer [12] constructed a subgraph of stars that served as a central communication hub to reduce the number of route hops between nodes. M. Fadhil, G. Owenson and M. Adda [18] proposed a clustering protocol for blockchain networks based on super nodes, called BCBSN (Bitcoin Clustering Based on Super Node). By clustering based on the locality of nodes, the propagation delay of transactions and blocks in the same cluster is reduced. The LBC (Location Based Clustering) protocol [19] and the BCBPT (Bitcoin Clustering Based on Ping Time) protocol [20] were further proposed. Nodes in the blockchain network were clustered according to physical location metrics such as their geographical location and Ping time to reduce the propagation delay of adjacent nodes.

Although these methods consider the impact of physical structure on performance, they do not completely solve the mismatch problem between the logical topology and the physical topology. Moreover, unlike the public blockchain, nodes in the consortium blockchain are more controllable, so these methods cannot optimally improve the performance of consortium blockchains.

## 3 PHYSICAL TOPOLOGY CONSTRUCTION METHODS

For structured P2P networks in consortium blockchains, we construct a matching physical topology among super nodes based on the multi-dimensional hypercube. Super nodes and ordinary nodes can be connected in a tree topology to achieve management, which is not described in detail here. In fact, the proposed topology applies not only to fully distributed P2P networks, but also to semi-distributed P2P networks.

In the original hypercube topology for a base $b = 2$, each node has the same responsibility. The network diameter, defined as the shortest path between most distant nodes in terms of node hops, is $\log_2 N$, which is less than $O(N)$ ($N$ is the number of nodes) [14]. As a result, the hypercube-based network topology has good characteristics in many aspects such as balanced load and good redundancy.

### 3.1 General Physical Topology

In the general blockchain network, we adopt the multi-dimensional hypercube or its variants to construct the physical topology. A complete hypercube is kind of a closed, compact and convex graph, whose 1-dimensional skeleton is composed of a group of line segments of equal length aligned to each dimension in the space where they are located, in which the relative line segments are parallel to each other, while the line segments intersecting at a point are orthogonal to each other. Fig. 1 shows examples of topologies constructed based on hypercubes in 3 to 5 dimensions, where crosses indicate that the corresponding nodes and links are unassigned or invalid in the network. Fig. 1. (a) is a complete 3-dimensional cube with 8 nodes and 12 links. Fig. 1. (b) is an incomplete 4-dimensional hypercube with 15 nodes and 29 links. Fig. 1. (c) is an incomplete 5-dimensional hypercube with 32 nodes and 75 links. The blue and the purple respectively represent the low-dimensional hypercube before and after the movement, and the green is the movement path.

For each node, an ID is assigned to uniquely identify the node. At this point, the hypercube topology supports the establishment of links between pairs of nodes at a distance of $2^i$ to improve query efficiency, that is, the logical distance between pairs of nodes whose IDs differ by only 1 bit is 1. Thus, the hypercube topology is a good match for upper protocols in the structured P2P network.

In the proposed physical topology, invalid links will be repaired actively in a finite time. If the P2P network is partitioned unfortunately, the isolated node will request data synchronization from its neighbors before resuming to work.

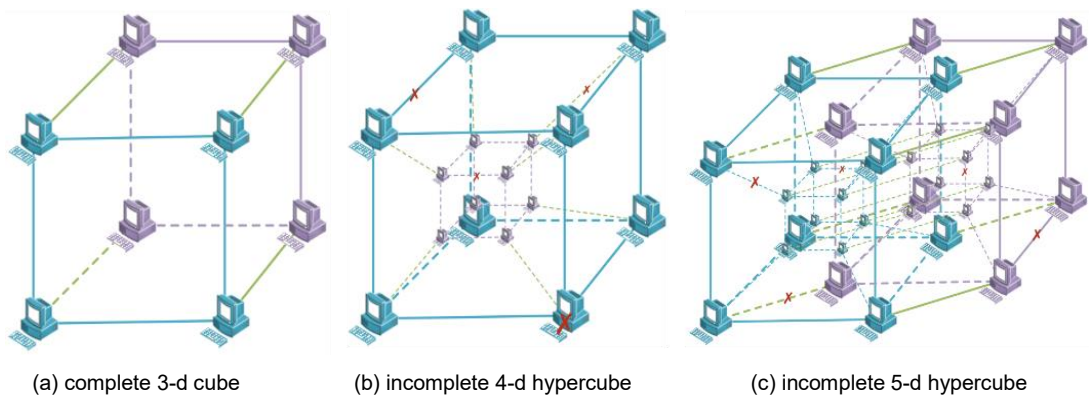

(a) complete 3-d cube　　(b) incomplete 4-d hypercube　　(c) incomplete 5-d hypercube

Fig. 1. Examples of topology construction with complete or incomplete hypercubes evolving from 3 to 5 dimension.

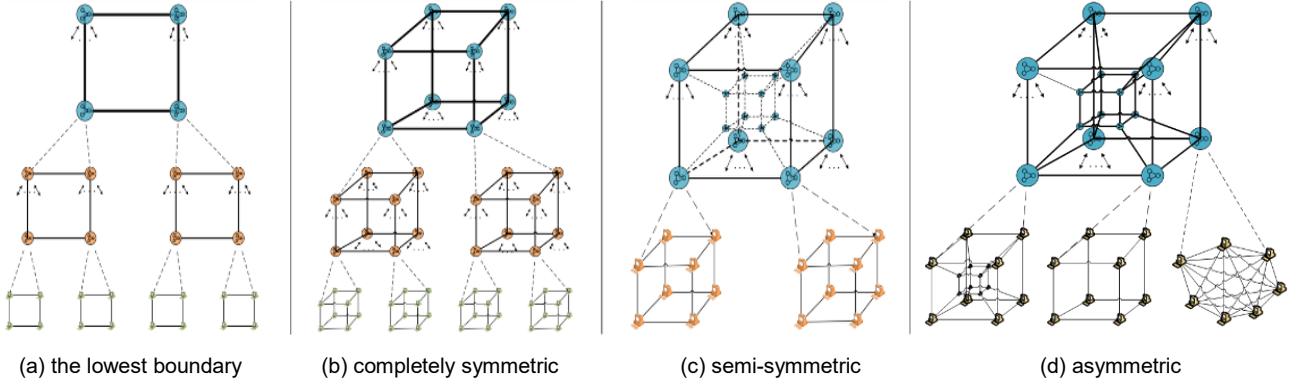

(a) the lowest boundary  (b) completely symmetric  (c) semi-symmetric  (d) asymmetric

Fig. 2. Examples of hierarchical recursive topology construction.

### 3.2 Hierarchical Recursive Physical Topology

With the expansion of network size, the use of the general hypercube topology will lead to excessive redundancy of links, and the complexity of upper-layer algorithms, such as routing algorithms, also increases. Therefore, recursion is a good method for scalability. In addition, the recursive topology should retain the advantage of strong regularity of hypercube.

In this section, we propose the hierarchical recursive topology, which is generated from the general hypercube-based topology in two steps, including recursion and interconnection. Each recursion turns the original node into a domain. Nodes in the same domain can construct a hypercube-based physical topology described in Section 3.1 or an arbitrary topology. The number of the domain at the $r$-th ($r \geq 2$) level is defined as $i_r$, which is equal to the smallest node number $n_{min}$ within it. After recursion, the new node number increases by one digit from the original node number, so the first $(r-1)$ digits of $i_r$ represent the recursive attribution of the domain from the first level to the $(r-1)$-th level. Assuming that the node number of the hypercube in the first-level domain $i_1 = 0$ is $\{n_0\} = \{0,1,2,\cdots,(2^{dim_0}-1)\}$, then the node number in the domain $i_r$ after $(r-1)$ recursions is $\{n_{i_r}\} = \{n_{i_{r-1}}\} \& \{0,1,2,\cdots,(2^{dim_{i_r}}-1)\}$, where & means splicing. The corresponding nodes of each domain at the $r$-th level are connected into the topological relationship of the domain at the $(r-1)$-th level with physical links, thus the interconnection between domains is completed.

Based on the above, we design three recursive methods to construct hierarchical physical topologies: completely symmetric, semi-symmetric, and asymmetric. In the completely symmetric way, each recursion takes the hypercube of the same dimension. In the semi-symmetric way, hypercubes of the same dimension are used in the same level, and that of different dimensions are used between levels. In the asymmetric way, each recursion takes a different dimensional hypercube or whatever. Fig. 2. (a) shows the lowest boundary situation of hierarchical recursive physical topology. Fig. 2. (b)-(d) show examples of the three above methods. Specifically, Fig. 2. (b) is a completely symmetric situation. Fig. 2. (c) is a semi-symmetric situation. Fig. 2. (d) is an asymmetric situation that there are 4, 8 and 4 domains with 4-dimensional hypercube, 3-dimensional hypercube and 7-potint full connection topology at the second level respectively.

We use Fig. 3 as the simplest example to describe the two steps to construct one completely symmetric topology of 2-dimensional hypercubes. The number in the circle indicates the node number, and the number with

$$\begin{aligned} p_{ji} &= P(X_j|X_i) \\ &= \sum_m P[\,(i-m)\ invalid\ links\ are\ repaired\,] \cdot P[\,(j-m)\ working\ links\ are\ invalid\,] \\ &= \sum_{m=max\{i+j-L,0\}}^{min\{i,j\}} \binom{i}{m} \mu^{i-m}(1-\mu)^m \cdot \binom{L-i}{j-m} \lambda^{j-m}(1-\lambda)^{L-i-j+m} \end{aligned} \qquad (1)$$

an underscore indicates the domain number. Suppose that the 4 nodes are numbered as $\{n_{i_1}\} = \{n_0\} = \{0,1,2,3\}$. In the first step, each node recurses into a 2-dimensional hypercube. There are 4 domains at the second level and 16 nodes, respectively numbered as $\{n_{i_2}\} = \{n_{00}, n_{10}, n_{20}, n_{30}\} = \{00,01,02,03\} \cup \{10,11,12,13\} \cup \{20,21,22,23\} \cup \{30,31,32,33\}$. In the second step, we connect the 4 nodes in the same domain to the nodes with the same last digit in the other two domains associated with them. The solid line represents the physical link between nodes, and the dashed line represents the logical link between domains. Specifically, the node 00 is linked to the nodes 10 and 30, the node 01 to the nodes 11 and 31, the node 02 to the nodes 12 and 32, the node 03 to the nodes 13 and 33, and so on.

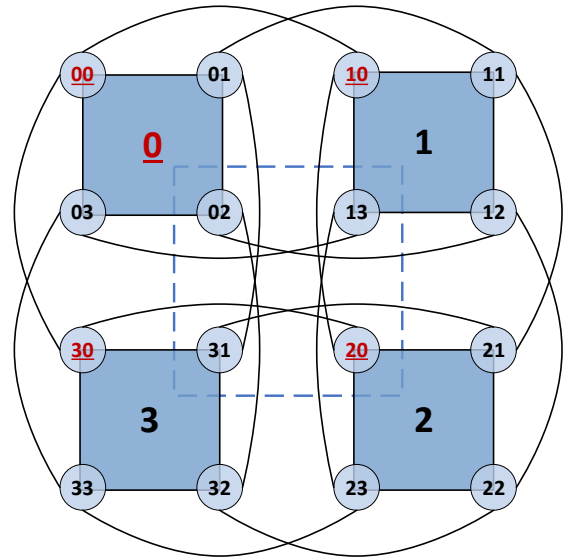

Fig. 3. Two steps to construct 1 completely symmetric topology of 2-dimensional hypercubes.

## 4 DISCUSSION

Within this section, we analyze the partition tolerance property of the proposed topology. We assume that the physical links between nodes are secure. Other assumptions of the topology construction and analyses are listed as follows.
1. Each involved link has only two states: work and failure.
2. Failure and repair of links are independent processes without memory.
3. MTBF (Mean Time Between Failures) and MTTR (Mean Time to Repair) of each link are independent, and their mean values are constant.
4. MTBF is much larger than MTTR.
5. If a partition has enough participating nodes, it is a good partition that works flawlessly. If not, it is a wrong partition.

Considering that the partition tolerance property reflects the reliability in blockchain, we propose two metrics: the partition tolerance probability and the average minimum repair time. The partition tolerance probability is defined as the probability that a good partition exists in the P2P network. The average minimum repair time is defined as the minimum time expected for the network to resume normal communication.

### 4.1 General Physical Topology

#### 4.1.1 Partition Tolerance Property
We first calculate the partition tolerance probability of the general physical topology. According to assumptions 1-5, since the state change of each link follows "work-failure-repair", a discrete-time Markov process can be utilized for mathematical modeling the partition tolerance problem [15]. As failure and repair processes are independent and memoryless, Fig. 4 shows the Markov chain of the hypercube-based topology consisting of $N$ nodes and $L$ links. The system state $X$ indicates the number of invalid links in the network.

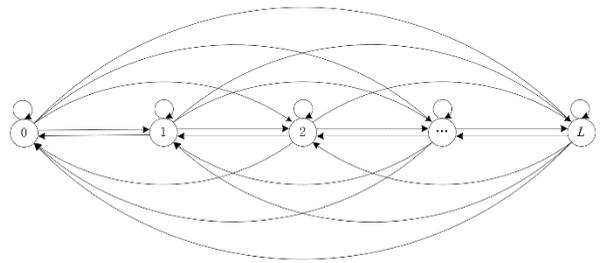

Fig. 4. The Markov chain of the general hypercube-based physical topology.

The transition matrix of the Markov model is denoted as $P_{(L+1)\times(L+1)}$, whose element $p_{ji}$ represents the probability of the system state transiting from $X_i$ to $X_j$. We calculate the state transition probability $p_{ji}$ in (1), where the intermediate variable $m \in [max\{i+j-L,0\}, min\{i,j\}]$ is the number of unrepaired invalid links during the transition process, $\lambda = 1/MTBF$ is the probability of a link to fail in the unit time, and $\mu = 1/MTTR$ is the probability of repairing it in the unit time.

Since Fig. 4 is a fully connected graph, the transition matrix $P$ satisfies randomness, irreducibility, and aperiodicity. According to the limit theorem of the Markov chain [16], the above Markov process eventually converges to a steady-state independent of the initial distribution. We obtain the steady-state probability vector $\pi = [\pi_0, \pi_1, \pi_2, \cdots, \pi_L]$ through the partitioning algorithm [17] with $O(L)$ iterations.

We estimate the partition tolerance probability by sampling. For each sample in steady-state $\pi_i$, the maximum number of nodes in the connected components is calculated to judge whether it is a wrong partition. The overall partition tolerance probability is

$$p = 1 - \sum_{i=1}^{L} \pi_i \cdot P\{\text{wrong partition}|\pi_i\}. \qquad (2)$$

Then we calculate the minimum repair time of the general physical topology. In the proposed topology, invalid links between big partitions should have priority to be repaired. Under this strategy, we define the minimum repair time as the time for the system to resume regular work. We get the overall average minimum repair time by taking the minimum repair time as the weight for each sample in (2).

### 4.1.2 Simulation

The simulation is based on the digital optical cable communication system which switches between primary and standby automatically. According to international standards, the communication systems (including physical links and repeaters) with the three different distances of $\{5000km, 3000km, 420km\}$ should meet the following indicators respectively

$$(MTBF, MTTR) \in \{(2190, 24), (3650, 14.4), (26070, 2.016)\}(h), \quad (3)$$

where $(h)$ is the unit and is short for $(hour)$.

So we take the pair of parameters as $(\lambda, \mu) \in \{(4.5662 \times 10^{-4}, 4.1667 \times 10^{-2}), (2.7397 \times 10^{-4}, 6.9444 \times 10^{-2}), (3.8358 \times 10^{-5}, 4.9603 \times 10^{-1})\} (/h)$. In the consortium blockchain, the minimum number of nodes in a good partition is not less than $k = \lfloor v/2 \rfloor + 1$. Fig. 5 compares the partition tolerance probability and average minimum repair time between the general multi-dimensional hypercube topology with the regular rooted tree topology and the ring lattice topology under the same degree.

The results show that with the increase of degree in the network, the partition tolerance probability increases rapidly without additional average minimum repair time. The proposed topology meets higher partition tolerance probability than the regular rooted tree topology and the ring lattice topology. In addition, the average minimum repair time for different topologies are similar, approximately equal to MTTR. Therefore, blockchains using the general hypercube-based topology can obtain excellent reliability in the partitioned network.

## 4.2 Hierarchical Recursive Physical Topology

### 4.2.1 Partition Tolerance Property

Denote the partition tolerance probability of a general topology at the $m$-th $(1 \leq m \leq r)$ level as $p_{i_m}$. The partition tolerance of the domain $i_r$ is not only affected by the topology it adopts at the $r$-th level but also related to the recursive path to which it belongs from the first level to the

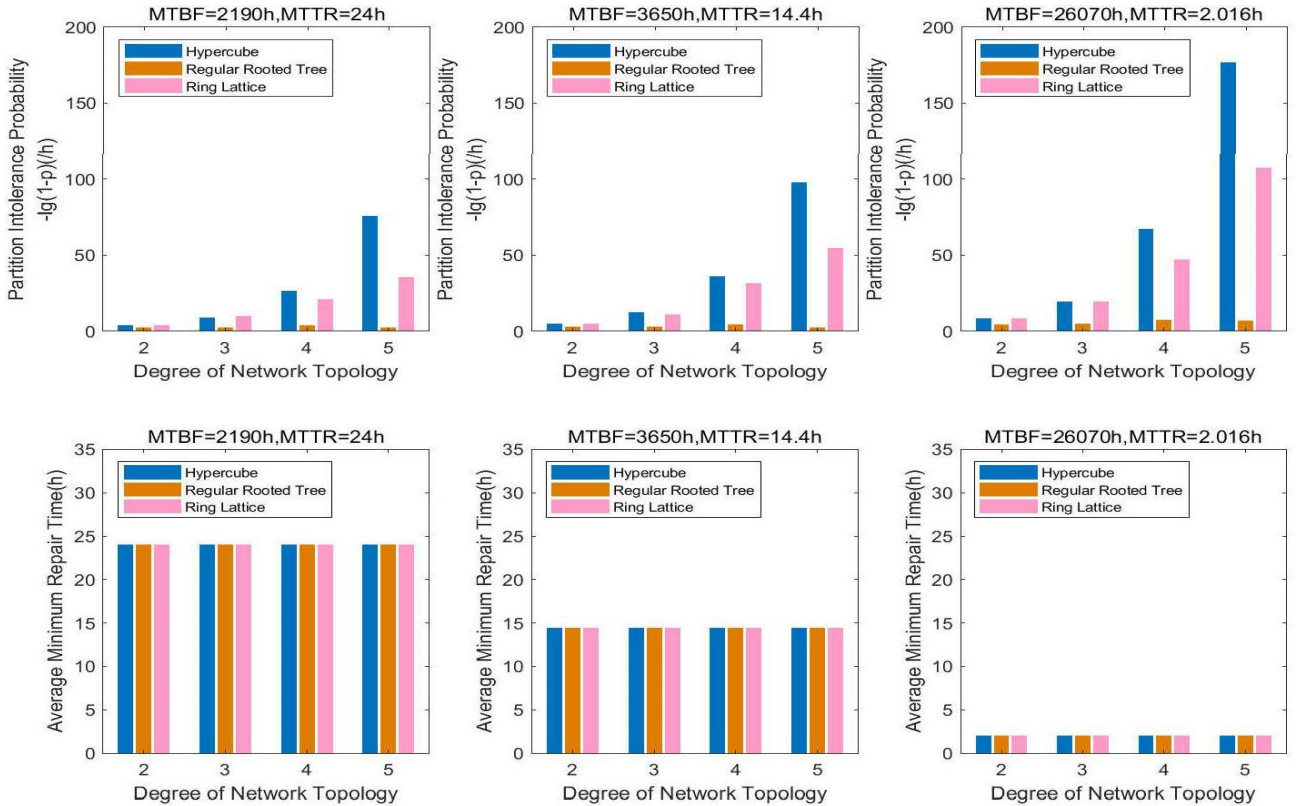

Fig. 5. Partition tolerance probability and average minimum repair time of the network topologies based on the general multi-dimensional hypercube, the regular rooted tree, and the ring lattice.

$(r-1)$-th level. Thus, the overall partition tolerance probability is

$$p = 1 - \sum_{m=1}^{r} \sum_{(i_1,i_2,i_3,\cdots,i_m)} p_{i_1} p_{i_2} p_{i_3} \cdots p_{i_{m-1}} (1 - p_{i_m}). \quad (4)$$

Similarly, let $t_{i_m}$ denote the average minimum repair time of a general topology at the $m$-th level, and the overall average minimum repair time is

$$t = \sum_{m=1}^{r} \sum_{(i_1,i_2,i_3,\cdots,i_m)} t_{i_m} \cdot \frac{p_{i_1} p_{i_2} p_{i_3} \cdots p_{i_{m-1}}(1-p_{i_m})}{1-p}. \quad (5)$$

According to (4) and (5), the recursive path from the first level to the $(r-1)$-th level has more influence on the overall partition tolerance than the topology of the domain itself at the $r$-th level. Therefore, a suggestion is that the topology with high partition tolerance probability and low average minimum repair time should be adopted in the upper recursion path.

### 4.2.2 Link Consumption

Considering that the number of nodes in the asymmetric recursive method is difficult to determine, in this section we only analyze the completely symmetric and the semi-symmetric recursive methods.

In a completely symmetric topology, we define the dimension of hypercubes in each domain as $dim_{i_m}$. Since each recursion takes the same dimension, for any $(i_1, i_2, i_3, \cdots, i_r)$, $dim_{i_m}$ is a fixed value and is denoted as $dim$. Obviously, the number of nodes for $(r-1)$ recursions is $N_{symm,r} = 2^{dim \times r}$. The links in the topology consist of relatively short intradomain links and relatively long interdomain links, that is,

$$\begin{cases} L_{symm,r} = 2^{dim-1} \times dim \times 2^{(r-1)\times dim} + L_{symm,(r-1)} \times 2^{dim} \\ L_{symm,1} = 2^{dim-1} \times dim \end{cases}. (6)$$

Equation (6) is equivalent to

$$\begin{cases} \frac{L_{symm,r}}{2^{r\times dim}} = \frac{dim}{2} + \frac{L_{symm,(r-1)}}{2^{(r-1)\times dim}} \\ L_{symm,1} = 2^{dim-1} \times dim \end{cases}.$$

Hence,

$$\frac{L_{symm,r}}{2^{r\times dim}} = \frac{L_{symm,1}}{2^{dim}} + \frac{dim}{2} \times (r-1) = \frac{r \times dim}{2}.$$

So, the total number of links is

$$L_{symm,r} = 2^{r\times dim-1} \times r \times dim. \quad (7)$$

Table1 illustrates the number of nodes and links of 0,

TABLE 1
THE NUMBER OF NODES AND LINKS IN THE COMPLETELY SYMMETRIC RECURSIVE PHYSICAL TOPOLOGY

| Number of recursions | Dimension of hypercubes | Number of nodes | Number of links |
|---|---|---|---|
| 0 | 2 | 4 | 4 |
|   | 3 | 8 | 12 |
|   | 4 | 16 | 32 |
|   | 5 | 32 | 80 |
| 1 | 2 | 16 | 32 |
|   | 3 | 64 | 192 |
|   | 4 | 256 | 1024 |
|   | 5 | 1024 | 5120 |
| 2 | 2 | 64 | 192 |
|   | 3 | 512 | 2304 |
|   | 4 | 4096 | 24576 |
|   | 5 | 32768 | 245760 |

1 and 2 completely symmetric recursions with 2- to 5-dimensional hypercubes respectively.

In a semi-symmetric topology, since domains in the same level use hypercubes of the same dimension, for any $i_m$, $dim_{i_m}$ is a fixed value. Similarly, the number of nodes for $(r-1)$ recursions is $N_{semi,r} = 2^{\sum_{m=1}^{r} dim_{i_m}}$, and the number of links satisfies

$$\begin{cases} L_{semi,r} = 2^{dim_{i_r}-1} \times dim_{i_r} \times 2^{\sum_{m=1}^{r-1} dim_{i_m}} + L_{semi,(r-1)} \times 2^{dim_{i_r}} \\ L_{semi,1} = 2^{dim_{i_1}-1} \times dim_{i_1} \end{cases}. (8)$$

So, the total number of links is

$$L_{semi,r} = 2^{\sum_{m=1}^{r} dim_{i_m}-1} \times \sum_{m=1}^{r} dim_{i_m}. \quad (9)$$

Table 2 illustrates the number of nodes and links of 0, 1 and 2 semi-symmetric recursions respectively with the recursive path of 4-, 3- and 2-dimensional hypercubes in order.

TABLE 2
THE NUMBER OF NODES AND LINKS IN THE SEMI-SYMMETRIC RECURSIVE PHYSICAL TOPOLOGY

| Number of recursions | Dimension of hypercubes | Number of nodes | Number of links |
|---|---|---|---|
| 0 | 4 | 16 | 32 |
| 1 | 4-3 | 128 | 448 |
| 2 | 4-3-2 | 512 | 2304 |

TABLE 3
COMPARISON OF LINK CONSUMPTION AND PARTITION TOLERANCE UNDER DIFFERENT PHYSICAL TOPOLOGY CONSTRUCTION METHODS

| Number of nodes | Topology construction method | | Number of links | | | Partition tolerance probability $-lg(1-p)(/h)$ | Average minimum repair time |
|---|---|---|---|---|---|---|---|
| | | | 5000km | 3000km | 420km | | |
| 64 | Regular rooted tree | | 63 | 0 | 0 | 2.79 | 24.0 |
| | Ring lattice | | 192 | 0 | 0 | 43.7 | 24.0 |
| | Hypercube | 0 recursion (6) | 192 | 0 | 0 | 87.0 | 24.0 |
| | | 1 completely symmetric recursion (3-3) | 96 | 96 | 0 | 8.62 | 24.0 |
| | | 2 completely symmetric recursions (2-2-2) | 64 | 64 | 64 | 3.56 | 21.4 |
| | | 1 Semi-symmetric recursions (4-2) | 128 | 64 | 0 | 3.53 | 14.4 |
| 4096 | Regular rooted tree | | 4095 | 0 | 0 | 3.93 | 24.0 |
| | Ring lattice | | 24576 | 0 | 0 | 107 | 24.0 |
| | Hypercube | 0 recursion (12) | 24576 | 0 | 0 | 227 | 24.0 |
| | | 1 completely symmetric recursion (6-6) | 12288 | 12288 | 0 | 87.0 | 24.0 |
| | | 2 completely symmetric recursions (4-4-4) | 8192 | 8192 | 8192 | 26.4 | 24.0 |
| | | 2 Semi-symmetric recursions (5-4-3) | 10240 | 8192 | 6144 | 16.6 | 2.02 |

Table 3 compares the properties of link consumption and partition tolerance under different physical topology construction methods. It is assumed that the 0th, 1st and 2nd recursion adopts links of 5000km, 3000km and 420km respectively, and the indicators in (3) in Section 4.1.2 are satisfied.

While each recursion does not reduce the total number of links, it actually uses more medium links or short links between nodes, greatly reducing the average minimum repair time. The higher the dimension of hypercubes in the recursion path, the lesser the repair time is required. On the other hand, although the hypercube-based topology has a higher number of links than the regular rooted tree topology, its partition tolerance property is much better and already meets the needs of actual consortium blockchains. Blockchain projects can freely choose the hierarchical recursive method according to its own reliability and cost requirements.

## 5 EXPERIMENTAL RESULTS

In this section, we present a systematic evaluation of the proposed physical topology on its performance to support the gossip protocol and the consensus protocol in the consortium blockchain.

### 5.1 Overlaying the Gossip Protocol

We build the P2P network with the general physical topology in the PeerSim-1.0.5 simulator [8]. For comparison, we also implement the regular rooted tree topology and the ring lattice topology. Upon the simulated physical layer, we execute a simple gossip protocol [9] and show its performance. In the gossip protocol, each node periodically selects a random subset of 4 neighbors from its cached set to initiate a message exchange operation. Since the out-degree is a fixed value equal to the cache size, we consider only the in-degree. Each simulation runs for 5000 cycles, and every cycle has 100 units of time.

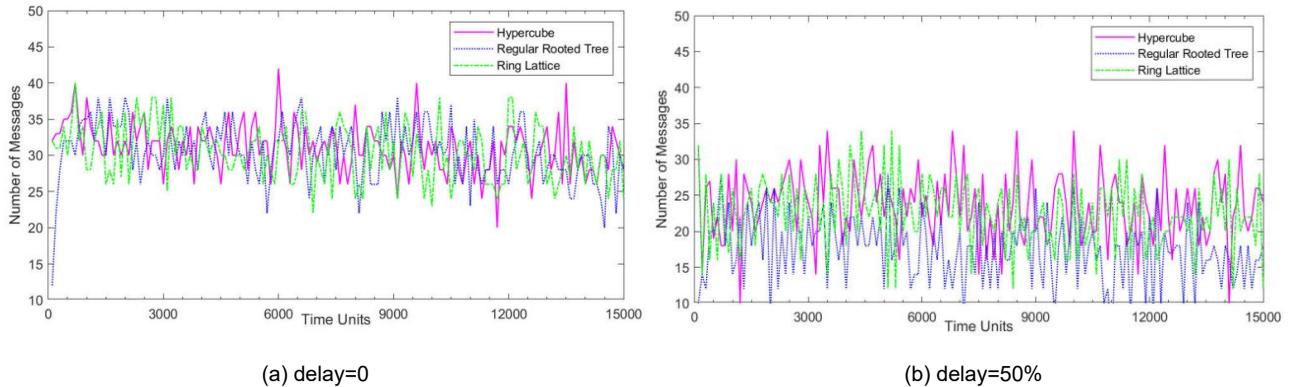

(a) delay=0

(b) delay=50%

Fig. 6. The number of gossip messages forwarded when N=16.

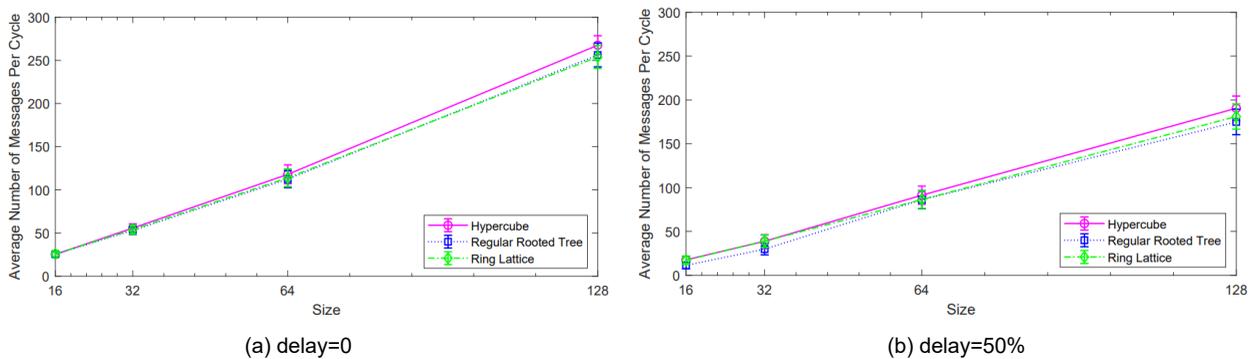

(a) delay=0

(b) delay=50%

Fig. 7. The average number of gossip messages forwarded with different number of nodes.

When the network size is $N = 16$, and the delay is 0 and 50%, the number of forwarded messages in the whole network is shown in Fig. 6. (a) and (b), respectively. In the ideal network with delay=0, the transmission performance of the three topologies compared is similar. In the network with delay=50%, the performance of the hypercube and the ring lattice topologies is similar and significantly better than that of the regular rooted tree topology.

As the network size increases, the average number of forwarded messages increases approximately linearly, as shown in Fig. 7. The experimental results show that for the used gossip protocol, the hypercube-based physical topology hardly brings additional forwarding redundancy.

We further prototype the hierarchical recursive physical topology in the PeerSim-1.0.5 simulator. For the four examples in Fig. 2 in Section 3.2, we test their performance separately, as shown in Fig. 8. It can be seen that the 50% delay leads to an approximately 50% reduction in the total number of messages forwarded, so the gossip protocol on the hierarchical recursive topology remains compliant with the original network characteristics. The results also show that recursion increases the number of forwarded messages by 10%-20% compared to the general hypercube-based topology. This is because the gossip protocol tends to cause loops in the domain when recursing, resulting in a large number of redundant messages in the network.

### 5.2 Overlaying the Consensus Protocol

We deploy the proposed physical topology across 8 servers in China, U.K., New Zealand and Malaysia. Each server has two 8-core CPUs and 10 Gbps network bandwidth. We next perform the PPoV (Parallel Proof of Vote) protocol [11] over different topologies. PPoV is a consortium consensus protocol with strong

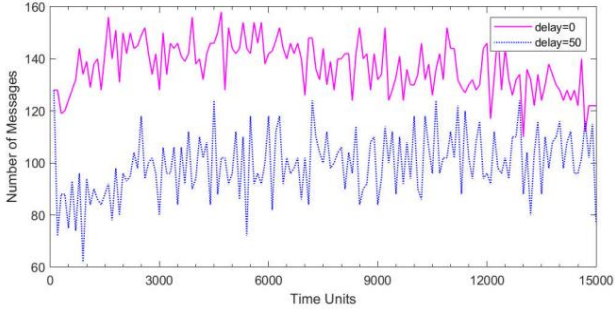
(a) lowest boundary topology (N=64)

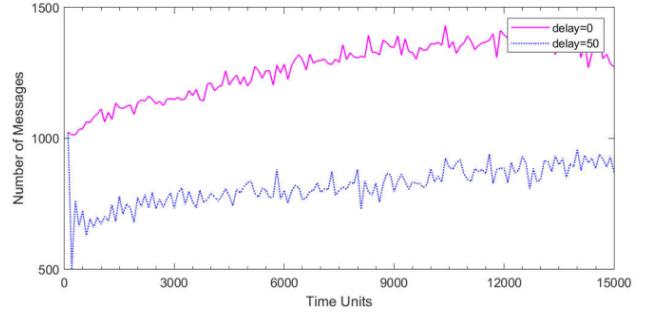
(b) completely symmetric topology (N=512)

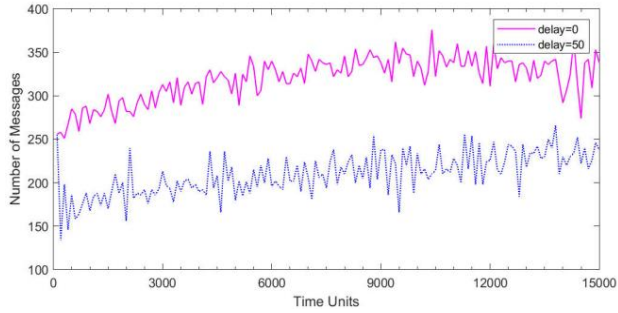
(c) semi-symmetric topology (N=128)

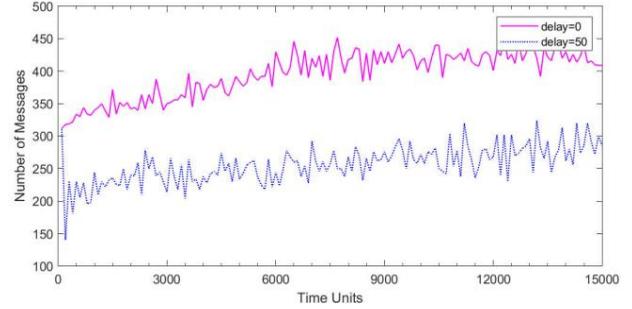
(d) asymmetric topology (N=156)

Fig. 8. The number of gossip messages forwarded with hierarchical recursive physical topologies.

consistency and high availability on a failure-free network. As a non-forked BFT consensus protocol in the consortium blockchain, any attempt by an attacker to change the consistency of PPoV consensus ultimately leads to a longer consensus time or even a timeout, so we use throughput as an indicator. Our experiment generates an average of 60,000 transactions per second, each with a size of 24 Bytes. A block can store up to 10,000 transactions, and its maximum size is 235MB.

Due to our resource limitation, we do not run any ordinary node. Because we are evaluating the impact of the physical topology on the consensus protocol only, we turn off signature verification and transaction execution to ensure enough computation resources for super nodes.

Fig. 9 compares the efficiency of the consensus protocol on the general hypercube topology and the star topology when the network size is $N = 4$. It can be seen that the throughput of the general hypercube topology is better than that of the star topology, and is approximately equal to the transaction generation rate.

In the above 4-node experiment, the PPoV consensus algorithm periodically selects a super node as the leader node randomly which occupies a high bandwidth. In the following large-scale experiment, in order to obtain relatively stable and believable throughput values, we

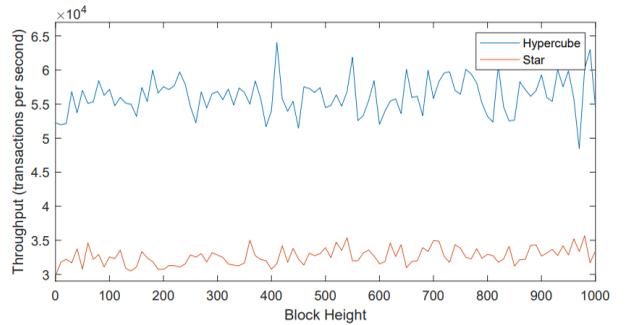
Fig. 9. The throughput of the consensus protocol when N=4 with the original leader selection mechanism.

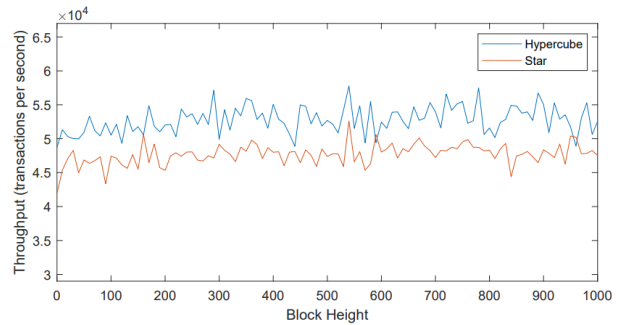
Fig. 10. The throughput of the consensus protocol when N=16 with the changed leader selection mechanism.

consider the best case that the central node of the star topology is always the leader node during the measurement period, and the rotation period is greater than 1000 consensus rounds. Fig. 10 compares the efficiency of the consensus protocol on the hierarchical recursive topology in Fig. 3 in Section 3.2 and the star topology when the network size is $N = 16$. Comparing Fig. 9 and Fig. 10, with the leader selection mechanism changed, the difference between the hypercube and the star topology is smaller for higher $N = 16$. However, the 1-recursive topology of the 2-dimensional hypercube still optimally supports the upper-layer consensus protocol.

According to observation during the experiment, under the above parameters, no matter how the number of nodes changes, each super node can achieve 100% utilization of a single CPU. In this case, the throughput is only affected by the network transmission rate. Fig. 11 shows the average throughput in networks of different sizes. It can be seen that compared with the star topology, the performance of the consensus protocol on top of the hypercube-based topology is more stable. In other words, the proposed physical topology not only does not affect the performance of upper-layer protocols, but also has good scalability.

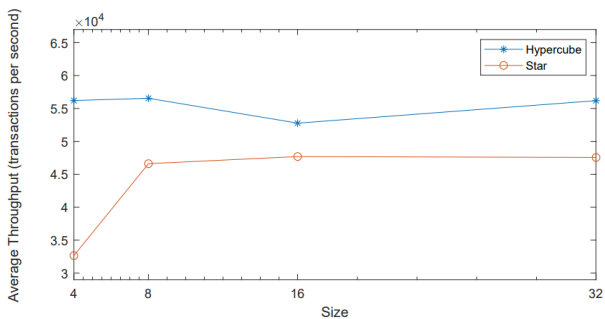

Fig. 11. The average throughput of the consensus protocol in the whole network.

## 6 CONCLUSION

In this paper, we propose a novel physical topology of the P2P network in blockchains based on the multi-dimensional hypercube to optimize the partition tolerance. We also extend the general topology to the hierarchical recursive topology for more medium links or short links. Through analyzing the partition tolerance by metrics of the partition tolerance probability and the average minimum repair time with the convergent Markov model and simulations on the digital optical cable communication system, we prove that the proposed topology meets better partition tolerance than the regular rooted tree topology and the ring lattice topology. Hierarchical recursion makes the hypercube topology satisfy excellent partition tolerance with less repair time. Experimental results from the prototype also show that since the proposed topology is at the physical layer, there is no need to modify upper-layer protocols. That is, blockchains constructed by the proposed topology can reach the CAP guarantee bound with proper transmission and consensus protocols meeting strong consistency and availability.

Our future work will focus on developing the logical communication protocol to solve the loop problem during data transmission in the hypercube-based physical topology, especially in the hierarchical recursive topology. We will also improve the analytical model by accurately and comprehensively extracting the feature parameters in simulation to achieve higher precision.

## ACKNOWLEDGMENT

This work was supported by Shenzhen Fundamental Research Program (No. GXWD20201231165807007-20200807164903001), GuangDong Prov. R&D Key Program (No. 2019B010137001), National Keystone R&D Program of China (No. 2017YFB0803204), Natural Science Foundation of China (NSFC) (No. 61671001), and Shenzhen Research Programs (No. JCYJ20190808155607340, No. JSGG20170406144032901, No. JSGG20170824095858416, No. JCYGJ20170306092030521).

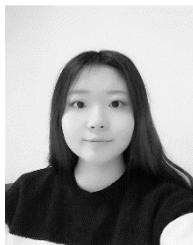
**Han Wang** received the B.Eng. degree from the Department of Communication Engineering, Jilin University of Technology, China, in 2017. She is currently pursuing the Ph.D. degree with the School of Information Science, Peking University, China. Her research interests include blockchain, distributed systems and cyber security.

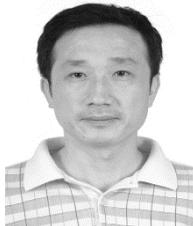
**Hui Li** is now a Professor of Peking University Shenzhen Graduate School. He received his B.Eng. and M.S. degrees from School of Information Eng., Tsinghua University, Beijing, China, in 1986 and 1989 respectively, and Ph.D. degree from the Dept. of Information Engineering, The Chinese University of Hong Kong in 2000. He was Director of Shenzhen Key Lab of Information theory & Future Internet architecture, Director of PKU Lab of CENI (China Environment for Network Innovations). He proposed the first co-governing future networking "MIN" Architecture in the world and this project "MIN: Co-Governing Multi-Identifier Network Architecture and Its Prototype on Operator's Network" was obtained the award of World Leading Internet Scientific and Technological Achievements by the 6th World Internet Conference on 2019, WuZhen, China. He was invited as Guest Editor of ZTE COMMUNICATIONS March 2020 Vol. 18 No. 1 with topic: Domain Name and Identifier of Internet: Architecture & Systems. The first English monograph by theme of "Cyberspace UN" has been published by Springer Publisher with title：
《Co-governed Sovereignty Network: Legal Basis and Its Prototype & Applications with MIN Architecture》. His research interests include network architecture, cyberspace security, distributed storage, and blockchain.

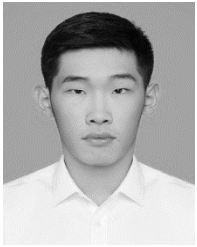

**Zixian Wang** received the B.Eng. degree from the Department of Software Engineering, Dalian University of Technology, China, in 2019. He is currently pursuing a M.S degree in School of Information Science, Peking University, China. His research interests include distributed system and future internet architecture.

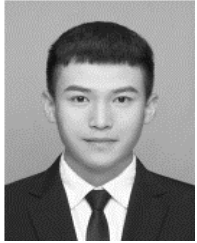

**Baofu Han** received the B.Eng. degree from the Electronic Information Engineering, Liaoning Shihua University, China, in 2018, and the the M.Eng. degree in Singla and Information Processing, Shenyang University of Technology, China, in 2021. He is currently a research assistant in School of Information Science, Peking University, China. His research interests include blockchain and distributed systems.

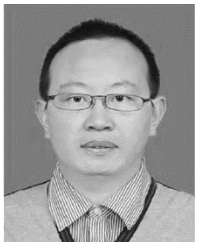

**Minglong Zhang** is a lecturer and postdoctoral research fellow with the Department of Electrical and Electronic Engineering, Auckland University of Technology, New Zealand. He received his MS degree and PhD from Peking University and Auckland University of Technology in 2011 and 2020, respectively. From 2011 to 2016, he was a researcher in The Chinese University of Hong Kong. His research interests include machine learning, 5G V2X communications, wireless communication and networks, as well as B5G/6G technologies.

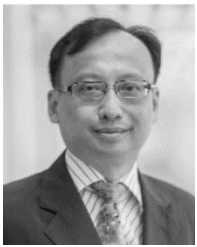

**Peter Han Joo Chong** (Senior Member, IEEE) received the Ph.D. degree from The University of British Columbia, Canada, in 2000. He was an Associate Professor (tenured) with the School of Electrical and Electronic Engineering, Nanyang Technological University (NTU), Singapore. From 2011 to 2013, he was an Assistant Head of the Division of Communication Engineering. From 2013 to 2016, he was the Director of INFINITUS, Centre for Infocomm Technology. He is currently a Professor and the Associate Head of School (Research), School of Engineering, Computer and Mathematical Sciences, Auckland University of Technology, Auckland, New Zealand. He is also an Adjunct Professor with the Department of Information Engineering, The Chinese University of Hong Kong, Hong Kong. His research interests are in the areas of mobile communications systems, including MANETs/VANETs, multi-hop cellular networks, and the Internet of Things/Vehicles.

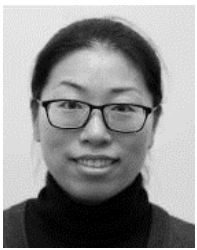

**Xiaoli Chu** (Senior Member, IEEE) received the B.Eng. degree in electronic and information engineering from Xi'an Jiao Tong University, in 2001, and the Ph.D. degree in electrical and electronic engineering from the Hong Kong University of Science and Technology, in 2005. From September 2005 to April 2012, she was with the Centre for Telecommunications Research, King's College London. Her research interest includes modeling, analysis, and algorithm design for improving the performance and efficiency of wireless communication systems.

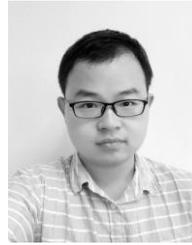

**Yang Liu** received the B.Eng. degree in microelectronics from Anhui University, Hefei, China, in 2012, the M.Eng. degree in software engineering from Beijing University of Aeronautics and Astronautics, in 2016, and the Ph.D. degree in communication and information system from Shanghai Institute of Microsystem and Information Technology (SIMIT), Chinese Academy of Sciences, Shanghai, in 2019. He is currently a research fellow with The University of Sheffield (UK). His current research interests include mobile computing, Internet of Things, and wireless localization.

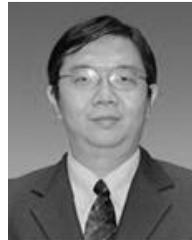

**Soung-Yue Liew** received his Bachelor's degree in Electrical Engineering from the National Taiwan University, Taiwan, in 1993, and the M.Phil. and Ph.D. degrees in Information Engineering from the Chinese University of Hong Kong (CUHK), Hong Kong, in 1996 and 1999, respectively. After graduation, he joined the Department of Information Engineering at CUHK as an Assistant Professor (1999–2000) and then as a Research Associate (2001–2002). From 2002 to 2003, he was a Research Associate at the Polytechnic University, New York (now known as New York University Tandon School of Engineering). In August 2003, he joined the Universiti Tunku Abdul Rahman, Selangor, Malaysia, at where he is currently an Associate Professor and the Dean of the Faculty of Information and Communication Technology. His research interest is in Internet of Things, Data Analytics, Mobile and Wireless Networking, Algorithm Design, Performance Analysis, etc.

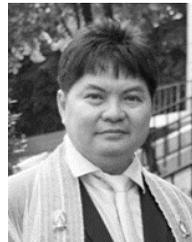

**Lunchakorn Wuttisittikulkij** received his Bachelor's degree in Electrical Engineering from Chulalongkorn University, Thailand, in 1990, and M.Sc. in Telecommunication and Information Systems and Ph.D. degrees from the University of Essex, the United Kingdom, in 1992 and 1997, respectively. After graduation, he joined the Department of Electrical Engineering at Chulalongkorn University as a Lecturer (1997–1999) and then as an Assistant Professor (1999–2002). From 2002, he becomes an Associate Professor. His research interest is in 5G and beyond wireless communication ecosystems.